\documentclass[12pt]{article}
\usepackage{latexsym}
\usepackage{amssymb}
\usepackage{amsmath}
\usepackage{graphicx}
\newcommand{\pme}{^{\prime}}
\newcommand{\gam}{\gamma(T,R)}
\newcommand{\alf}{\alpha(T,R)}
\newcommand{\lam}{\lambda(r,t)}
\newcommand{\nw}{\nu(r,t)}
\newtheorem{theorem}{Theorem:}

\begin{document}

\title{\bf{\small The $T$-Domain and Extreme Matter
Phases Inside Spherically Symmetric Black Holes}}
\author{{\small A. DeBenedictis}
\footnote{Department of Physics, Simon Fraser University, Burnaby,
British Columbia, Canada, V5A 1S6, e-mail: adebenea@sfu.ca,
adebened@langara.bc.ca} \and {\small D. Aruliah} \footnote{Institute of
Applied Mathematics, University of British Columbia, Vancouver, British
Columbia, Canada, V6T 1Z2} \and {\small A. Das} \footnote{Department
of Mathematics, Simon Fraser University, Burnaby, British Columbia,
Canada, V5A 1S6, e-mail: das@sfu.ca}}
\date{{\small August 30, 2001}}

\maketitle

\newpage
\begin{abstract}
Black hole interiors (the $T$-domain) are studied here in great detail.
Both the {\em general} and particular $T$-domain
solutions are presented including non-singular ones. Infinitely many
local $T$-domain solutions may be modeled with this scheme. The duality
between the $T$ and $R$ domains is presented. It is
demonstrated how generally well behaved $R$-domain solutions will give
rise to exotic phases of matter when collapsed inside the event horizon.
However, as seen by an external observer, the field is simply that of the
Schwarzschild vacuum with well behaved mass term and no evidence of
this behaviour may be observed. A singularity theorem is also
presented which is independent of energy conditions.
\end{abstract}

\vspace{1.6mm}

\section{Introduction}
\qquad Over the years there have been extensive studies carried out in the
literature on the subject of static, spherically symmetric stars. These
regular solutions consist of some spherical matter field smoothly patched
to the Schwarzschild vacuum solution at some surface outside the
gravitational radius or event horizon. Many of these solutions involve
bodies composed of perfect fluids \cite{ref:kramer}, some of which employ
equations of state to supplement the equations of relativity.

\qquad Specific models of spherically symmetric gravitational collapse
have also been of much interest since the pioneering work of
Oppenheimer and Snyder \cite{ref:opsnyd}. A collection of work
regarding specific collapse models may be found in \cite{ref:kramer}
and other standard references. Interestingly, some studies of collapsing
spherically
symmetric stars consisting of anisotropic fluid matter reveal a
transition into exotic matter after collapsing past their event
horizons (\cite{ref:das1}, \cite{ref:das2}). Black holes of this type
are known in the literature as exotic black holes.

\qquad Black hole interiors which are not singular at the classical
level (regular black holes) have also been studied
\cite{ref:mars}-\cite{AyonGarcia99b}. Although it is believed that any
serious candidate theory for quantum gravity must eliminate the
singularity, it is instructive to study what criteria must be met at
the classical level to remove this singularity. We present such
solutions here as well as derive the general properties non-singular
solutions must possess.

\qquad For the above reasons, it is interesting to study the $T$-domain in
more detail, especially addressing the issue of extreme exotic matter. It
may be argued that the black hole evaporation process \cite{ref:hawkevap}
tends to reveal more of the black hole interior as time progresses.
Therefore, studies of the interior region have relevance to exterior
observers in this way. The $T$-domain also provides an interesting arena
in
which to perform theoretical studies of matter under extreme gravitational
conditions.

\qquad This paper serves several purposes. First we present an
investigation of the general $T$-domain. Although much work has
been
done regarding spherical systems in the $R$-domain, relatively little
analytic work has been performed on the $T$-domain. We believe it is
useful
to present the general local solution to the $T$-domain problem as it
sheds light on some of the origins of the exotic matter phases
mentioned above. It turns out that the presence of an apparent horizon
induces drastic phase transitions in the matter field. Also, it is
hoped that the general solution will provide a useful starting point
for future research in this area.

\qquad In the literature, analytic studies in this domain involve
specific solutions which depend on the $T$ coordinate only, so called
$T$-spheres which are eternal black holes. For examples of such studies
the reader is referred to the interesting work by Ruban
\cite{ref:ruban1} \cite{ref:ruban2} and Poisson and Israel
\cite{ref:poiss} as well as references above. However, physically, a
$T$-domain solution is expected to result from the gravitational
collapse of an $R$-domain system (some well behaved star). In other
words, before the formation of an event horizon, the collapse is
described by a metric depending on the exterior radial and time
coordinate ($r$ and $t$ respectively). Therefore, after the system has
collapsed within its Schwarzschild radius forming a black hole, the
corresponding interior solution should also depend on both of these
coordinates (labeled $T$ and $R$ respectively in the $T$-domain). One
particular solution constructed here does possess such dependence.

\qquad The interesting particular solutions which are
presented include the $T$-domain analogue of the constant density star.
It is found that the corresponding matter in the $T$-domain possesses
substantially different physical properties as will be discussed later
in the paper. Finally, we solve the $T$-domain equations to construct
non-singular solutions, one of which depends on both the radial and time
coordinate. The other regular  solution may represent the late stages of
collapse of a polytropic star which does {\em not} form a singularity.

\qquad This study places some emphasis on the presence of exotic matter
and the elimination of the central singularity is not a requirement of
all solutions. These extreme matter phases are considered and studied
in some detail as they are motivated by the collapse studies mentioned
earlier and it is hoped that the analysis here will shed some light on
the origin of such phases. This leads us to another motivation for
this study: It is interesting to note that, when considering solutions
which are well behaved in the $R$-domain, their extension into the
$T$-domain is usually accompanied by {\em exotic} physical properties.
We show that ordinary astrophysical systems can naturally form exotic
systems after collapsing in a black hole. One simple example we
demonstrate, in the general solution, is that demanding a well behaved
Schwarzschild mass measured by an observer in the $R$-domain requires
material which possesses a local {\em tension}.

\section{Physics in the $T$-domain}
\qquad We begin here with a brief analysis of the $R$-domain and present
the
general solution in this domain. This is useful as it will later be
compared to the general $T$-domain solution which possesses substantially
different physical properties even though, at first glance, the
$T$-coordinate chart may seem very similar to the $R$-chart except for
signs.
The solution
presented here allows for the study of any collapse model in the
$T$-domain.

\qquad The spherically symmetric line element in curvature coordinates is
furnished by:
\begin{equation}
ds^{2}=-e^{\nu(r,t)}\,dt^{2}+e^{\lambda(r,t)}\,dr^{2} +r^{2}\,d\theta^{2}
+r^{2}\sin^{2}\theta\,d\phi^{2}; \label{eq:one}
\end{equation}
\begin{equation}
r_{1}<r<r_{2},\;\; t_{1} < t < t_{2},\;\;\; 0<\theta <\pi, \;\;\; 0\leq
\phi <2\pi. \nonumber
\end{equation}
The geometry is governed by the Einstein field equations
\footnote{Conventions here follow those of \cite{ref:MTW} with $G=c=1$.
The Riemann tensor is therefore given by $R^{\mu}_{\;\rho\nu\sigma}=
\Gamma^{\mu}_{\;\rho\sigma,\nu}+
\Gamma^{\mu}_{\;\alpha\nu}\Gamma^{\alpha}_{\;\rho\sigma} -\ldots$ with
$R_{\rho\sigma}=R^{\alpha}_{\;\rho\alpha\sigma}$.}
\begin{equation}
R_{\mu\nu}-\frac{1}{2}R\,g_{\mu\nu}=8\pi T_{\mu\nu}, \label{eq:einst}
\end{equation}
supplemented with the conservation law
\begin{equation}
T^{\mu\nu}_{\;\;\; ;\nu}=0. \label{eq:conslaw}
\end{equation}
Explicitly, in mixed form, (\ref{eq:einst}) yields the following
non-trivial equations:
\begin{subequations}
\begin{align}
-8\pi
T^{t}_{\;t}=&\frac{1}{r^{2}}+\frac{e^{-\lam}}{r}\left(\lam_{,r}-\frac{1}{r}
\right)\; , \label{eq:twoa} \\
-8\pi T^{r}_{\;r}=&
\frac{1}{r^{2}}-\frac{e^{-\lam}}{r}\left(\nw_{,r}+\frac{1}{r}\right)
\; ,
\label{eq:twob} \\
-8\pi T^{r}_{\;t}=&\frac{\left(e^{-\lam}\right)_{,t}}{r} \; ,
\label{eq:twoc}
\\
-8\pi T^{\theta}_{\;\theta} =& -8\pi
T^{\phi}_{\;\phi} = \frac{e^{-\nw}}{2}
\left[\lambda_{,t,t}+\frac{1}{2}\left(\lam_{,t}^{\;\;2}-
\lam_{,t}\nw_{,t}\right)\right] \nonumber \\
&-\frac{e^{-\lam}}{2}\left[\nw_{,r,r}+\frac{1}{2}
\left(\nw_{,t}^{2}-\nw_{,r}\lam_{,r}\right)\right. \nonumber
\\
&\left.+\frac{1}{r}
(\nw-\lam)_{,r}\right]. \label{eq:twod}
\end{align}
\end{subequations}
The conservation law, (\ref{eq:conslaw}), yields only two non-trivial
equations
\begin{subequations}
\begin{align}
T^{r}_{\;r,r}+ T^{t}_{\;r,t} +\left(\frac{1}{2}
\nw_{,r}+\frac{2}{r}\right)T^{r}_{\;r}+\frac{1}{2}(\lam+\nw)_{,t}
T^{t}_{\;r}& \nonumber \\
-\left[\frac{1}{2}\nw_{,r}T^{t}_{\;t} +\frac{2}{r}
T^{\theta}_{\;\theta}\right]=&0 , \label{eq:threea} \\
T^{t}_{\;t,t}+T^{r}_{\;t,r}+\frac{1}{2} \lam_{,t}\left(
T^{t}_{\;t}-T^{r}_{\;r}\right)+
\left[\frac{1}{2}(\lam+\nw)_{,r}+\frac{2}{r}\right]T^{r}_{\;t}=&0.
\label{eq:threeb}
\end{align}
\end{subequations}
In the system of six partial differential equations,
(\ref{eq:twoa}-\ref{eq:twod}) and (\ref{eq:threea}-b), there exist six
unknown functions: $\lam$, $\nw$, and the four components of the
stress-energy
tensor. However, there are two differential identities among the equations
and therefore, one can either prescribe two functions to make the system
determinate or determine them from other means.

Synge's strategy \cite{ref:syngebook} of solving the
equations (\ref{eq:twoa}-\ref{eq:threeb}) is the following:
\begin{itemize}
\item Prescribe $T^{t}_{\;t}$ and solve the equation (\ref{eq:twoa}) for
$\lam$.

\item Prescribe $T^{r}_{\;r}$ (which may be related to $T^{t}_{\;t}$ by an
equation of state) and solve a linear combination of
(\ref{eq:twoa}) and (\ref{eq:twob}) for $\nw$.

\item Define $T^{r}_{\;t}$ by equation (\ref{eq:twoc}).

\item Define $T^{\theta}_{\;\theta}$ by the conservation equation
(\ref{eq:threea}).

\item At this stage, by the differential identities, one can show that
{\em all} equations are satisfied.
\end{itemize}

Following this strategy, the {\em most general solution} of the system
of equations in a suitable domain $D_{r}$ of the $r$-$t$ plane can be
furnished as:
\begin{subequations}
\begin{align}
e^{\lam}=&1+\frac{8\pi}{r}\int_{r_{0}}^{r} T^{t}_{\;t}(r\pme,t) r^{\prime
2}\,dr\pme-\frac{f(t)}{r}=: 1-\frac{2m(r,t)}{r},  \label{eq:foura} \\
e^{\nw}=& \left[1-\frac{2m(r,t)}{r}\right]
\exp\left\{h(t)+8\pi\int_{r_{0}}^{r}\left[\frac{T^{r}_{\;r}(r\pme,t)
-T^{t}_{\; t}(r\pme,t)}{r\pme-2m(r\pme,t)}\right]r^{\prime 2}\,
dr\pme\right\}, \label{eq:fourb}\\
T^{r}_{\;t}=&\frac{1}{4\pi r^{2}}m(r,t)_{,t}\;\; , \label{eq:fourc} \\
T^{\theta}_{\;\theta}\equiv& T^{\phi}_{\;\phi} :=\left(\frac{r}{2}\right)
\left(T^{r}_{\;r,r}+ T^{t}_{\;r,t}\right)
+\left(1+\frac{r}{4}\nw_{,r}\right)T^{r}_{\;r} \nonumber \\
&+\frac{r}{4}\left(\lam+\nw\right)_{,t}T^{t}_{\;r} -\frac{r}{4}
\nw_{,r}T^{t}_{\;t}. \label{eq:fourd}
\end{align}
\end{subequations}
The functions $f(t)$ and $h(t)$ are arbitrary or free functions of
integration. By a coordinate transformation
$\hat{t}=\int\exp[h(t)/2]dt$, and subsequently dropping the hat
notation, one may rewrite (\ref{eq:fourb}) as:
\begin{equation}
e^{\nw}=\left[1-\frac{2m(r,t)}{r}\right] \exp\left\{
8\pi\int_{r_{0}}^{r}\left[\frac{T^{r}_{\;r}(r\pme,t)
-T^{t}_{\; t}(r\pme,t)}{r\pme-2m(r\pme,t)}\right]r^{\prime 2}\,
dr\pme\right\}. \label{eq:five}
\end{equation}
Usually, to avoid a singularity at $r=0$, $f(t)\equiv0$. However, we
shall retain it for future use.

\qquad The range of the $r$-coordinate in (\ref{eq:one}), prompted by the
general solution and energy conditions in (\ref{eq:foura}-d), and the
domain
$D_{r}$ are given by:
\begin{eqnarray}
0 &<&2m(r,t) <r_{1} \leq r_{0} < r <r_{2} \label{eq:six} \\
D_{r}&:=&\left\{(r,t) \in \mathbf{R}^{2}:\;r_{1} < r < r_{2},\;
t_{1}<t<t_{2}\right\}. \nonumber
\end{eqnarray}

\qquad In many problems the components of $T^{\mu}_{\;\nu}$
are due to specific fluids or fields. In these cases the number of
unknowns versus the number independent equations may be different than
suggested. However, the general solution presented above still
contains these (which may be variationally derived and determinate) as
special cases.

\qquad Otherwise, one is free to prescribe the energy density and radial
pressure. This method is useful in examinations of relativistic stellar
structure where one usually prescribes a reasonable energy and pressure
from nuclear theory and studies of plasmas \cite{ref:weinberg}.

\qquad We now turn our attention to the $T$-domain. Since this domain is
physically quite different from the previous, we adopt a new, hopefully
clear notation for quantities in this domain to avoid confusion. The
metric for the $T$-domain is furnished by
\begin{equation}
ds^{2}=-e^{\gam}\,dT^{2} + e^{\alf}\,dR^{2} +T^{2}\,d\theta^{2} +
T^{2}\sin^{2}\theta\,d\phi^{2}, \label{eq:seven}
\end{equation}
with
\begin{equation}
T_{1}<T<T_{2}, \;\; R_{1} < R < R_{2}, \;\; 0 < \theta <\pi, \;\; 0\leq
\phi < 2\pi. \nonumber
\end{equation}
The vacuum $T$-domain is given by the well known metric:
\begin{equation}
ds^{2}=-\left[\frac{2M}{T}-1\right]^{-1}\,dT^{2}+
\left[\frac{2M}{T}-1\right]\,dR^{2}+T^{2}\,d\theta^{2} +
T^{2}\sin^{2}\theta\,
d\phi^{2}. \label{eq:schwT}
\end{equation}
\qquad Einstein's field equations, $G^{\mu}_{\;\nu}=8\pi
\Theta^{\mu}_{\;\nu}$ yield:
\begin{subequations}
\begin{align}
-8\pi\Theta^{T}_{\;T}=&\frac{1}{T^{2}}+
\frac{e^{-\gam}}{T}\left(\alf_{,T}+\frac{1}{T}\right), \label{eq:eighta}
\\
-8\pi\Theta^{R}_{\;R}=&\frac{1}{T^{2}}-
\frac{e^{-\gam}}{T}\left(\gam_{,T}-\frac{1}{T}\right), \label{eq:eightb}
\\
-8\pi\Theta^{T}_{\;R}=&\frac{\left(e^{-\gam}\right)_{,R}}{T},
\label{eq:eightc} \\
-8\pi\Theta^{\theta}_{\;\theta}\equiv& -8\pi\Theta^{\phi}_{\;\phi}=
\frac{e^{-\gam}}{2}\left[\alf_{,T,T} +\frac{1}{2}\left(
\alf_{,T}^{\;\;\;2}
\right.\right. \nonumber \\
-&\left.\left.\alf_{,T}\gam_{,T}\right)
+\frac{1}{T}\left(\alf-\gam\right)_{,T}\right]
\nonumber \\
-&\frac{e^{-\alf}}{2}\left[\gam_{,R,R}
+\frac{1}{2}\left(\gam_{,R}^{\;\;\;2}
-\gam_{,R}\alf_{,R}\right)\right]. \label{eq:eightd}
\end{align}
\end{subequations}
The conservation equations, $\Theta^{\mu}_{\;\nu ;\mu}=0$ lead to
\begin{subequations}
\begin{align}
\Theta^{T}_{\;T,T}+&\Theta^{R}_{\;T,R}
+\frac{1}{2}\left(\alf+\gam\right)_{,R}\Theta^{R}_{\;T}
-\frac{1}{2}\alf_{,T}\Theta^{R}_{\; R}-
\frac{2}{T}\Theta^{\theta}_{\;\theta}
\nonumber \\
+&\left(\frac{1}{2}\alf_{,T}+\frac{2}{T}\right)\Theta^{T}_{\; T}=0,
\label{eq:ninea} \\
\Theta^{R}_{\;R,R}+&\Theta^{T}_{\;R,T}
+\left[\frac{1}{2}\left(\alf+\gam\right)_{,T}
+\frac{2}{T}\right]\Theta^{T}_{\;R} \nonumber \\
+&\frac{1}{2}\gam_{,R}\left(\Theta^{R}_{\; R} - \Theta^{T}_{\;T}\right)
=0. \label{eq:nineb}
\end{align}
\end{subequations}

\qquad The general solution of the system (\ref{eq:eighta}-d) and
(\ref{eq:ninea}-b) is given by:
\begin{subequations}
\begin{align}
e^{-\gam}=&\frac{1}{T} \left[\phi(R)- 8\pi \int_{T_{0}}^{T} T^{\prime 2}
\Theta^{R}_{\;R}(T\pme ,R)\,dT\pme\right]-1 =:\frac{2\mu (T,R)}{T}-1,
\label{eq:tena} \\
e^{\alf}=&
\exp\left\{\beta(R)+
8\pi\int_{T_{0}}^{T}\left[\frac{\Theta^{T}_{\;T}(T\pme,R)
-\Theta^{R}_{\;R}(T\pme,R)}{2\mu(T\pme,R)-T\pme}\right] T^{\prime2}
\,dT\pme\right\}  \nonumber \\
\times&\left[\frac{2\mu (T,R)}{T}-1\right],
\label{eq:tenb} \\
\Theta^{T}_{\;R}:=&\frac{1}{4\pi T^{2}} \mu(T,R)_{,R}, \label{eq:tenc} \\
\Theta^{\theta}_{\;\theta}\equiv& \Theta^{\phi}_{\;\phi}:=\frac{T}{2}
\left(\Theta^{T}_{\;T,T}+\Theta^{R}_{\;T,R}\right)
+\left[1+\frac{T}{4}\alf_{,T}\right]\Theta^{T}_{\;T} \nonumber \\
&+\frac{T}{4}\left(\alf+\gam\right)_{,R}\Theta^{R}_{\;T} -\frac{T}{4}
\alf_{,T} \Theta^{R}_{\;R}. \label{eq:tend}
\end{align}
\end{subequations}
Here, the functions $\phi(R)$ and $\beta(R)$ are arbitrary or free
functions of integration. The arbitrary function $\beta(R)$ may be
eliminated by a similar transformation as in the $R$-domain:
\begin{equation}
\hat{R}=\int\exp[\beta(R)/2]\,dR .
\end{equation}
Subsequently, dropping the hats, (\ref{eq:tenb}) yields
\begin{equation}
e^{\alf}=\left[\frac{2\mu (T,R)}{T}-1\right] \exp\left\{8\pi
\int_{T_{0}}^{T}\left[\frac{\Theta^{T}_{\;T}(T\pme,R)
-\Theta^{R}_{\;R}(T\pme,R)}{2\mu(T\pme,R)-T\pme}\right] T^{\prime 2}
\,dT\pme\right\}. \label{eq:eleven}
\end{equation}

\qquad A valid $T$-domain for the above solution is provided by
\begin{eqnarray}
0 &<& T_{1} \leq T_{0} < T_{2} < 2\mu(R,T) \nonumber \\
D_{T}&:=&\left\{(T,R)\in \mathbf{R}^{2}:\;\; T_{1} < T < T_{2}, \;\; R_{1}
< R < R_{2} \right\}. \label{eq:twelve}
\end{eqnarray}

\qquad It is evident that there exist unmistakable similarities between
the solution given in (\ref{eq:foura}-d) and the solution
(\ref{eq:tena}-d). These solutions, though similar, yield {\em
completely different physics} as will be discussed later. These
differences hinge on the fact that these are two completely different
charts leading to different physical quantities in each domain. For
example, (\ref{eq:tena}) shows that, to make a positive contribution
to the Schwarzschild mass, a negative pressure or tension is required.
There is, however a duality between the solutions in the two domains
which we will state in the form of a theorem.
\begin{theorem}
Let a set of solutions of the spherically symmetric gravitational
equations and of differentiability class $C^{3}$ in the domain $D_{r}$
of equation (\ref{eq:six}) be furnished by the equations in
(\ref{eq:foura}-d). Then, a distinct set of solutions in the domain
$D_{T}$ of equation (\ref{eq:twelve}) as provided in the equations
(\ref{eq:tena}-d) by the duality transformation symbolically denoted
as:
\begin{eqnarray}
r\rightarrow T, \; t\rightarrow R, \; \gamma=\lambda, \;
\alpha&=&\nu,\;
\phi=f ,\; \beta=h, \nonumber \\
\Theta^{R}_{\;R}&:=&T^{t}_{\;t}, \nonumber \\
\Theta^{T}_{\;T}&:=&T^{r}_{\;r}, \label{eq:thirteen} \\
\Theta^{T}_{\;R}&:=&T^{r}_{\;t}. \nonumber
\end{eqnarray}
\end{theorem}
Proof emerges directly from equations (\ref{eq:foura}-d) and
(\ref{eq:tena}-d). However, the duality transformation from {\em
domain} $D_{r}$ into {\em domain} $D_{t}$ does {\em not} exist. In
case, solutions in $D_{r}$ and $D_{T}$ represent two distinct
coordinate systems in the same universe, the corresponding coordinate
neighbourhoods are necessarily {\em disjoint}. Moreover, the domains
$D_{r}$ and $D_{T}$ may correspond to two distinct universes. It is
interesting to note that physically reasonable solutions in $D_{r}$
with, for example, $T^{t}_{\;t}(r,t) <0$ yield, by (\ref{eq:thirteen}),
necessarily exotic fluid in $D_{T}$ with $\Theta^{R}_{\;R}(T,R) < 0$.
Other applications of the above will be furnished later when we analyse
specific solutions.

\qquad It is useful at this point to compute the components of the
orthonormal Riemann tensor components in $D_{T}$ as this will later aid
our study of the singularity structure of the solution. We choose the
natural orthonormal basis from (\ref{eq:seven}) as:
\begin{equation}
e^{\mu}_{T}=e^{-\gamma /2}\delta^{\mu}_{T},
\;\;e^{\mu}_{R}=e^{-\alpha /2}\delta^{\mu}_{R}, \;\;
e^{\mu}_{\theta}=T^{-1}\delta^{\mu}_{\theta}, \;\;
e^{\mu}_{\phi}=(T\sin\theta)^{-1} \delta^{\mu}_{\phi}.
\end{equation}
In this frame, the curvature tensor posesses the following components as
well as
those related by symmetry:
\begin{subequations}
\begin{align}
R_{(TRTR)}=&\frac{1}{2} \left\{ e^{-\gam}\left[ \frac{1}{2}\gam_{,T}
\alf_{,T} -\frac{1}{2}(\alf_{,T})^{2} \right.\right. \nonumber
\\
&-\alf_{,T,T}\Bigr] +
e^{-\alf}\left[\gam_{,R,R}+\frac{1}{2}(\gam_{,R})^{2}
\right. \nonumber \\
&-\left.\left.\frac{1}{2}\gam_{,R}\,\alf_{,R}\right]\right\},
\label{eq:Rtrtr} \\
R_{(T\theta T\theta)}=&\frac{\gam_{,T}}{2T}e^{-\gam},
\label{eq:Rtthetattheta} \\
R_{(R\theta R\theta)}=&\frac{\alf_{,T}}{2T}e^{-\gam},
\label{eq:Rrthetartheta} \\
R_{(\theta\phi\theta\phi)}=&\frac{1+e^{-\gam}}{T^{2}}=\frac{2\mu(T,R)}{T^3},
\label{eq:Rthetaphithetaphi} \\
R_{(T\theta R\theta)}=&\frac{\gam_{,R}}{2T}
e^{-\frac{1}{2}\left(\gam+\alf\right)}.
\label{eq:Rtthetartheta}
\end{align}
\end{subequations}
Here, indices in parentheses are used to denote expressions calculated
in the orthonormal frame.

\qquad The singularity theorems of Penrose and Hawking \cite{ref:pensing},
\cite{ref:hawkpensing} are well known. These are proved under the
satisfaction of certain energy conditions. However, solutions
(\ref{eq:tena}-d)
also produce a singularity in $D_{T}$, without the introduction of energy
conditions, as will be demonstrated next.
\begin{theorem}
Let the metric functions $\alf$ and $\gam$ be at least of class $C^{3}$
and the stress-energy tensor be of class $C^{1}$ in the $T$-domain.
Moreover, let the tension function, $\mu(T,R)$, be of class $C^{3}$
and satisfy the inequality $\frac{2\mu(T,R)}{T}>1$ in $D_{T}$. In that
case, $\lim_{T\rightarrow 0_{+}}R_{(\theta\phi\theta\phi)}$ diverges
for all $R\in(R_{1},R_{2})$.
\end{theorem}
{\em Proof}: By the inequality $2\mu(T,R)/T > 1$ and the condition of
thrice
differentiability, it can be concluded that in the positive neighbourhood
of $T=0$,
\begin{equation}
\frac{2\mu(T,R)}{T}= 1+\left[H(T,R)\right]^{2}. \nonumber
\end{equation}
Here, $H(T,R)\neq 0$ is some function of class $C^{3}$. Denoting by
\begin{equation}
h(R):=\lim_{T\rightarrow 0^{+}} H(T,R)\;\; , \nonumber
\end{equation}
it is derived that
\begin{equation}
\lim_{T\rightarrow 0^{+}}\left[\frac{2\mu(T,R)}{T}\right]=1+[h(R)]^{2},
\;\; R\in(R_{1},R_{2}) . \nonumber
\end{equation}
By the above equation and (\ref{eq:Rthetaphithetaphi}) it is proved that
\begin{equation}
\lim_{T\rightarrow 0^{+}}R_{(\theta\phi\theta\phi)} = \lim_{T\rightarrow
0^{+}} \frac{1+[H(T,R)]^{2}}{T^{2}} =
\left\{1+[h(R)]^{2}\right\}\lim_{T\rightarrow
0^{+}}\frac{1}{T^{2}} \rightarrow \infty. \nonumber
\end{equation}
Thus, the conclusion of the stated theorem is proved. $\blacksquare$

This theorem has interesting consequences. Essentially, it states that
a spherically symmetric space-time {\em can not} possess a Lorentzian
$T$-domain metric which is regular at $T=0$ regardless of energy
conditions. If the space-time is to contain a black hole {\em and} be
regular at $T=0$ one must either abandon an everywhere Lorentzian
metric or else an ``inner'' Cauchy horizon must exist at some $T =
T_{i}$ such that $0 < T_{i} < T_{b} < 2M$, with $T_{b}$ being the
matter-vacuum boundary. This yields an $R$-domain type metric in the
region $0 < T < T_{i}$. This last fact has been previously noted in
\cite{ref:mars} using geodesic completeness. However, it is useful to
illustrate how this comes about from our local analysis. In light of the
no-hair theorem, the above theorem should apply to many non-spinning,
collapsing bodies even if they initially deviate from spherical symmetry.

\subsection{Patching to the Vacuum Solution}
\qquad In this section we address the issue of patching the interior
matter solution to the vacuum Schwarzschild line element given by
(\ref{eq:schwT}). At the junction, $T=B(R)$, the condition of Synge
\cite{ref:syngebook} is chosen:
\begin{equation}
\Theta^{\mu}_{\;\nu}\,n_{\mu\;\; |T=B(R)}=0, \label{eq:syngecond}
\end{equation}
where $n_{\mu}$ are the covariant components of a unit normal vector to
the
boundary $T=B(R)$. This boundary may be defined by a level curve of the
function $F(T,R):=B(R)-T=0$. We may use the gradient, $F(T,R)_{,\mu}$,
of this function to define components of the normal. Namely,
$n_{T}\propto F(T,R)_{,T}=-1$ and $n_{R}\propto F(T,R)_{,R}=B(R)_{,R}$
with other
components zero. The explicit junction conditions then read:
\begin{subequations}
\begin{align}
\left[\Theta^{R}_{\;R}B(R)_{,R}-\Theta^{T}_{\;R}\right]_{|T=B(R)}=0,
\label{eq:jun1} \\
\left[\Theta^{R}_{\;T}B(R)_{,R}-\Theta^{T}_{\;T}\right]_{|T=B(R)}=0.
\label{eq:jun2}
\end{align}
\end{subequations}
Solutions which cannot meet these conditions may still be useful as
local solutions.

\qquad Outside the matter region, $2M > T > B(R)$, the solution for
$g_{TT}$ yields, via (\ref{eq:tena}),
\begin{equation}
g_{TT}=-\left(\frac{2\mu(B(R),R)}{T}-1\right)^{-1}. \label{eq:gttbound}
\end{equation}
The function $\mu(B(R),R)$ is the total invariant Schwarzschild mass.
Assuming for the moment that the junction conditions are met, the fact
that $\mu(B(R),R)$ is indeed a constant may be shown utilizing
(\ref{eq:tenc}) along with the condition (\ref{eq:jun1}):
\begin{equation}
2\mu(B(R),R)=f(R)-
8\pi\int_{T_{0}}^{B(R)} \Theta^{R}_{\;R}(T\pme,R)T^{'2}dT\pme.
\end{equation}
Therefore,
\begin{eqnarray}
2\mu(B(R),R)_{,R}&=& 2\left\{\mu(T,R)_{,R}|_{T=B(R)} +
\mu(T,R)_{,T}|_{T=B(R)} B(R)_{,R} \right\} \nonumber \\
&=&-8\pi [B(R)]^{2}\left[ \Theta^{R}_{\; R}(T,R) B(R)_{,R} -
\Theta^{T}_{\;R}(T,R)\right]|_{T=B(R)}  \nonumber \\
&=& 0.
\end{eqnarray}
When this is patched to the vacuum Schwarzschild $T$-domain metric
(\ref{eq:schwT}) it is clear that the parameter $\mu(B(R),R)=M$ is
indeed identical to the Schwarzschild mass as observed by an external
observer. The very interesting point is the following: Observers
outside the black hole (in the $R$-domain) feel the usual effects of
gravity for a Schwarzschild black hole or star of mass $M$. The
gravitational effects, however, are more likely generated by a {\em
tension} rather than an energy density. This tension generated mass
leads us to the term ``exotic matter''.

\qquad At the boundary, the non-vacuum solution becomes:
\begin{eqnarray}
ds^{2}|_{B_{-}(R)}=&-&\left[\frac{2M}{B(R)}-1\right]^{-1}\,dT^{2}+
\left[\frac{2M}{B(R)}-1\right]
\exp\left[S(R)\right]\,dR^{2} \nonumber \\
&+&[B(R)]^{2}\,d\theta^{2} +
[B(R)\sin\theta ]^{2}\,d\phi^{2}, \label{eq:dsbound}
\end{eqnarray}
with $S(R)$ given by:
\begin{equation}
S(R):=8\pi\int_{T_{0}}^{B(R)}T^{\prime2}
\left[\frac{\Theta^{R}_{\;R}(T\pme,R)-
\Theta^{T}_{\;T}(T\pme,R)}{T\pme+8\pi\int_{T_{0}}^{T\pme}
T^{\prime\prime}
\Theta^{R}_{\;R}(T^{\prime\prime},R)\;dT^{\prime\prime}}\right]
dT\pme. \label{eq:S}
\end{equation}
The $\exp[S(R)]$ term may be absorbed into the definition of a new radial
coordinate, $\hat{R}$, so that (\ref{eq:dsbound}) becomes:
\begin{eqnarray}
ds^{2}|_{\hat{B}_{-}(\hat{R})}=
&-&\left[\frac{2M}{\hat{B}(\hat{R})}-1\right]^{-1}\,dT^{2}+
\left[\frac{2M}{\hat{B}(\hat{R})}-1\right]
d\hat{R}^{2} \nonumber \\
&+&[\hat{B}(\hat{R})]^{2}\,d\theta^{2} +
[\hat{B}(\hat{R})\sin\theta ]^{2}\,d\phi^{2}.
\end{eqnarray}
Dropping hats subsequently, it may be seen that this is indeed the limit
$T\rightarrow B_{+}(R)$ of the vacuum black hole solution
(\ref{eq:schwT}).

\qquad We shall now briefly discuss the Israel \cite{ref:israelcond}
junction condition in the $T$-domain which demands continuity of the
second
fundamental form at the stellar boundary. The extrinsic curvature tensor
of the non-null hypersurface $T=B(R)$ is given by
\begin{subequations}
\begin{align}
K_{RR}=&\frac{1}{\sqrt{\left|e^{-\alf} (B\pme(R))^{2}
-e^{-\gam}\right|}} \left\{B^{\prime\prime}(R) \right. \nonumber \\
&+\frac{1}{2}\left[ e^{\alf-\gam}\alf_{,T}
+\left(2\gam
-\alf\right)_{,R}B\pme(R)\right. \nonumber \\
&+\left(\gam-2\alf\right)_{,T}(B\pme(R))^{2}
\nonumber \\
&\left.\left.
-e^{\gam-\alf}\gam_{,R}(B\pme(R))^{3}\right]_{|T=B(R)}\right\}, \\
K_{\theta\theta}=
&\frac{B(R)e^{-\gamma(B(R),R)}}{\sqrt{|e^{-\alf}
(B\pme(R))^{2}-e^{-\gam}|}}, \\
K_{\phi\phi}=&\sin^{2}\theta\,K_{\theta\theta}.
\end{align}
\end{subequations}
This completes the analysis of the $T$-domain general solution.

\subsection{Eternal Black Holes}
\qquad Consider now the special, spatially homogeneous cases, where
there is an additional Killing vector given by
$\frac{\partial}{\partial R}$. This is the $T$-domain analogue of static
black holes. Following the previous prescription, the solution is
furnished by:
\begin{subequations}
\begin{align}
e^{-\gam}=&\frac{1}{T} \left[A- 8\pi \int_{T_{0}}^{T} T^{\prime 2}
\Theta^{R}_{\;R}(T\pme)\,dT\pme\right]-1 =:\frac{2\mu (T)}{T}-1,
\label{eq:etena} \\
e^{\alf}=& \exp\left\{K+
8\pi\int_{T_{0}}^{T}\left[\frac{\Theta^{T}_{\;T}(T\pme)
-\Theta^{R}_{\;R}(T\pme)}{2\mu(T\pme)-T\pme}\right] T^{\prime2}
\,dT\pme\right\}  \nonumber \\
\times&\left[\frac{2\mu(T)}{T}-1\right],
\label{eq:etenb} \\
\Theta^{T}_{\;R}\equiv& 0, \label{eq:etenc} \\
\Theta^{\theta}_{\;\theta}\equiv \Theta^{\phi}_{\;\phi}:=&\frac{T}{2}
\Theta^{T}_{\;T,T}
+\left[1+\frac{T}{4}\alpha(T)_{,T}\right]\Theta^{T}_{\;T} \nonumber \\
& -\frac{T}{4} \alpha(T)_{,T} \Theta^{R}_{\;R}. \label{eq:etend}
\end{align}
\end{subequations}
Here, $A$ and $K$ are constants of integration.

\qquad The boundary hypersurface is given by the simple equation:
\begin{eqnarray}
T&=&B(R):=T_{b}=\hbox{a constant}, \\
B\pme(R)&=& B^{\prime\prime}(R)\equiv 0. \nonumber
\end{eqnarray}
The junction conditions of Synge (\ref{eq:jun1}, \ref{eq:jun2}) reduce
to
\begin{subequations}
\begin{align}
\Theta^{T}_{\;R}(T_{b})\equiv& 0 \label{eq:eternalbc1} \\
\Theta^{T}_{\;T}(T_{b})=& 0. \label{eq:eternalbc2}
\end{align}
\end{subequations}
The first junction condition is automatically satisfied via
(\ref{eq:etenc}) so that the matter field must only obey
(\ref{eq:eternalbc2}).

\qquad The Israel conditions boil down to the continuity of the metric
as well as the following extrinsic curvature components:
\begin{subequations}
\begin{align}
K_{RR}(T_{b})=&\frac{1}{2}
e^{-\gamma(T_{b}) /2}(e^{\alpha(T)})_{,T\;_|T=T_{b}}\;\;, \\
K_{\theta\theta}=&T_{b}e^{-\gamma(T_{b})/2}.
\end{align}
\end{subequations}
We next consider both spatially homogeneous and non-homogeneous
particular solutions.

\section{Particular Solutions}
\qquad Here we investigate particular solutions of the general case
described above. We first consider a special, tension generated
solution provided by:
\begin{subequations}
\begin{align}
\Theta^{T}_{\;T}(T)\equiv& 0, \\
\Theta^{R}_{\;R}(T)=&-\frac{1}{8\pi T^{2}}\,\delta(T) <0 ,
\end{align}
\end{subequations}
with $\delta(T)$ the Dirac delta function. In this case, the equations
(\ref{eq:etena},b) yield the $T$-domain Schwarzschild metric of
(\ref{eq:schwT}). (A rescaling of the $R$-coordinate is tacitly assumed.)

\qquad It is interesting to note that the source of this field
may be interpreted as possessing a velocity which is of a tachyonic
nature.
The above stress-energy tensor may be written as:
\begin{equation}
\Theta^{\mu}_{\;\nu}=-\frac{1}{8\pi T^{2}}\delta(T)s^{\mu}s_{\nu}
\nonumber
\end{equation}
with
\begin{equation}
s^{T}=s^{\theta}=s^{\phi}\equiv 0, \;\;\; s^{\alpha}s_{\alpha}=+1.
\nonumber
\end{equation}
The eigenvalue equation, $\Theta^{\mu}_{\;\nu}s^{\nu}=-1/(8\pi
T^{2})\delta(T) s^{\mu}$, indicates that $s^{\mu}$ may be
interpreted as a space-like dynamical mean velocity of the source. This is
also evident from visual inspection of the standard Kruskal-Szekeres
diagram.

\qquad Next we study the well known interior solution of Schwarzschild
\cite{ref:schwint}. In the $R$-domain it is furnished by:
\begin{eqnarray}
ds^{2}&=&-\left[\frac{3\sqrt{|1-qr_{b}^{2}|} -
\sqrt{|1-qr^{2}|}}{3\sqrt{|1-qr_{b}^{2}|}-1}\right]^{2}\,dt^{2}+
\left[1-qr^{2}\right]^{-1}\,dr^{2} \nonumber \\
& &+r^{2}\,d\theta^{2} + r^{2}\sin^{2}\theta\,d\phi^{2} , \nonumber \\
T^{r}_{\;r}&=&T^{\theta}_{\;\theta}=T^{\phi}_{\;\phi} =\frac{3q}{8\pi}
\left[\frac{\sqrt{|1-qr^{2}|}-\sqrt{|1-qr_{b}^{2}|}}{3
\sqrt{|1-qr_{b}^{2}|}-\sqrt{|1-qr^{2}|}}\right] > 0, \label{eq:sintsol} \\
T^{t}_{\;t}&=&-\frac{3q}{8\pi} < 0, \nonumber \\
T^{r}_{\; t} &\equiv& 0, \nonumber \\
T^{r}_{\; r}(r_{b})&=&0, \nonumber \\
0&<& r \;\;<\;\; r_{b} \;\; < \;\; \frac{1}{\sqrt{q}}. \nonumber
\end{eqnarray}
Here $q > 0$ is a constant proportional to the mass density and
$r=r_{b} >0$ is the boundary of the spherical star. We now apply the
duality to the above solution and show that its $T$-domain analogue
possesses quite distinct physical properties.

\qquad In the $T$-domain, the solution is a function of the corresponding
time-like coordinate only. This arises from the ``static'' condition
which  is a common simplification in studies of $R$-domain stellar
structure exact solutions. In the $R$-domain, the assumption that the
radial
pressure, $T^{r}_{\;r}$, is equal to the transverse pressure now
becomes:
\begin{equation}
\Theta^{T}_{\;T}=\Theta^{\theta}_{\;\theta}\equiv
\Theta^{\phi}_{\;\phi}.
\end{equation}
These conditions give rise to the solution:
\begin{eqnarray}
ds^{2}&=& -\left(qT^{2}-1\right)^{-1}\,dT^{2} +
\left[\frac{3\sqrt{|qT_{b}^{2}-1|}-\sqrt{\left|qT^{2}-1\right|}}{3
\sqrt{|qT^{2}_{b}-1|}-1}\right]^{2}\,dR^{2} \nonumber \\
&+& T^{2}\, d\theta^{2} + T^{2}\sin^{2}\theta\, d\phi^{2}, \nonumber \\
\Theta^{R}_{\;R}&=&-3\frac{q}{8\pi} < 0, \label{eq:partic} \\
\Theta^{T}_{\;T}&=&\Theta^{\theta}_{\;\theta}\equiv
\Theta^{\phi}_{\;\phi}=-3\frac{q}{8\pi}\left[\frac{\sqrt{|qT^{2}_{b}-1|}-\sqrt{\left|qT^{2}-
1\right|}}{3\sqrt{|qT^{2}_{b}-1|}-\sqrt{\left|qT^{2}-1\right|}}\right]
<0, \Theta^{T}_{\;T}(T_{b})=0, \nonumber \\
0&<& T_{i} \;\;<\;\;T\;\;<\;\;T_{b}. \nonumber
\end{eqnarray}

\qquad We now investigate the energy conditions regarding the above
solution. The eigenvalue problem, $\Theta^{\mu}_{\;\alpha}v^{\alpha}=
\lambda v^{\mu}$, is trivial since $\Theta^{\mu}_{\;\nu}$ is diagonal.
The eigenvectors, $v^{\mu}$, define an orthonormal tetrad:
$v^{\mu}_{(T)}$, $v^{\mu}_{(R)}$, $v^{\mu}_{(\theta)}$ and
$v^{\mu}_{(\phi)}$ (the corresponding eigenvalues denoted by
$\lambda_{(T)}$, $\lambda_{(R)}$,$\lambda_{(\theta)}$ and
$\lambda_{(\phi)}$). The mixed stress-energy momentum tensor admits a
decomposition in terms of its eigenvalues and eigenvectors. This
decomposition is
\begin{equation}
\Theta^{\mu}_{\;\nu}=
\Theta^{(\alpha)}_{\;\;(\beta)}v^{\mu}_{(\alpha)}v^{(\beta)}_{\nu}.
\label{eq:thetamunudecomp}
\end{equation}
With the notation
\begin{equation}
p_{\perp}:=\lambda_{(\theta)}=\lambda_{(\phi)}=\lambda_{(T)}:=-\rho=
-\frac{3q}{8\pi}\left[\frac{(qT_{b}^{2}-1)^{1/2}
-(qT^{2}-1)^{1/2}}{3(qT_{b}^{2}-1)^{1/2}-(qT^{2}-1)^{1/2}}\right] < 0
\nonumber
\end{equation}
and
\begin{equation}
\lambda_{(R)}:=p_{\shortparallel}=-\frac{3q}{8\pi}< 0, \nonumber
\end{equation}
the equation (\ref{eq:thetamunudecomp}) allows us to write
\begin{equation}
\Theta^{\mu}_{\;\nu}=\left(\rho-|p_{\shortparallel}|\right)s^{\mu}s_{\nu}-
\rho\delta^{\mu}_{\;\nu}. \label{eq:innertmunu}
\end{equation}
Here $s^{\mu}=v^{\mu}_{(R)}$, $s^{\mu}s_{\mu}=1$ and
$s^{\theta}=s^{\phi}=s^{T}\equiv 0$. This is somewhat analogous to the
case for a perfect fluid except that the isotropic pressure is
replaced by a radial {\em tension} and the velocity of the fluid is
{\em space-like}. This makes the fluid {\em tachyonic} in nature
rather than a perfect fluid. The fluid is anisotropic since the
angular tensions, $p_{\perp}$, are equal in magnitude to $\rho$ which
differs from the radial tension, $p_{\shortparallel}$. Although the
energy density is positive, the weak, strong and dominant energy
conditions are not satisfied since
\begin{subequations}
\begin{align}
\rho+p_{\perp}=&0, \label{eq:violation1} \\
\rho+p_{\shortparallel}=&\frac{-6q(qT^{2}_{b}-1)^{1/2}}{8\pi
\left[3(qT_{b}^{2}-1)^{1/2}-(qT^{2}-1)^{1/2}\right]} < 0.
\label{eq:violation2}
\end{align}
\end{subequations}
Thus, the fluid matter is {\em exotic} matter \footnote{In the
literature, exotic matter usually violates energy conditions because
the energy density $\rho < 0$ which is not the case here. However,
another common feature in studies of exotic matter is that the
principal stresses are tensions rather than pressures. For this
reason, the matter is still called exotic.}. This solution constitutes
an eternal exotic black hole because the exotic matter lies entirely
within the $T$-domain. Observers in domains $D_{I}$ of the
Kruskal-Szekeres spacetime see a black hole of Schwarzschild mass
$M=qT^{3}_{b}/2$ even though the $T$-domain is substantially different.

\qquad At the stellar boundary, $T=T_{b}$, we wish to employ the
Israel boundary condition. The components of the extrinsic curvature
tensor for the interior solution at $T_{b}$ are given by:
\begin{subequations}
\begin{align}
K_{RR|T_{b}^{-}}=&-2\frac{qT_{b}\sqrt{qT_{b}^{2}-
1}}{\left[3\sqrt{qT_{b}^{2}-1}-1\right]^{2}}, \label{eq:krrspec} \\
K_{\theta\theta|T_{b}^{-}}=&T_{b}\sqrt{qT^{2}_{b}-1}.
\label{eq:kththspec}
\end{align}
\end{subequations}
Those of the $T$-domain vacuum solution (\ref{eq:schwT}) are given by:
\begin{subequations}
\begin{align}
K_{RR|T_{b}^{-}}=&-\frac{M}{T_{b}^{2}}\sqrt{\left(\frac{2M}{T_{b}}-1\right)}\exp[S],
\label{eq:krrvac} \\
K_{\theta\theta|T_{b}^{-}}=&T_{b}\sqrt{\left(\frac{2M}{T_{b}}-1\right)},
\label{eq:kththvac}
\end{align}
\end{subequations}
referring to (\ref{eq:dsbound}) and (\ref{eq:S}) for the definition of
$S$. Metric continuity at $T_{b}$ along with the result
$M=\frac{q}{2}T_{b}^{3}$, immediately gives
\begin{equation}
\exp[S]=\frac{4}{T_{b}\left(3\sqrt{\frac{2M}{T_{b}}-1}-1\right)^{2}},
\nonumber
\end{equation}
and therefore the extrinsic curvature and metric are continuous at
$T=T_{b}$. The solution is shown schematically in figure 1.
\begin{figure}[ht!]
\begin{center}
\includegraphics[bb=0 0 565 525, scale=0.4,
keepaspectratio=true]{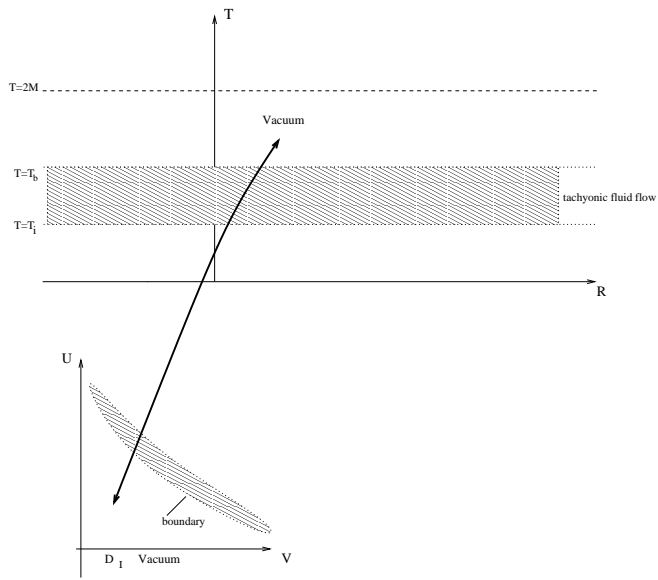} \caption{{\small Qualitative
representation of exotic matter existing in the $T$-domain along with
the image in the Kruskal spacetime.}} \label{fig1}
\end{center}
\end{figure}

\qquad To complete the above analysis, consider the tension generated
Schwarz\-schild mass, $M$, as a constant and the parameter $q$ as a
variable. In terms of $M$ and $q$, set the boundary parameters as
\begin{equation}
T_{i}:=q^{-1/2} \;\;\;\hbox{and}\;\;\;
T_{b}:=\left(\frac{2M}{q}\right)^{1/3}.
\end{equation}
Hence, $D_{I}$, is as wide as it possibly can be since $T_{i}$ is as
small as it can be for a prescribed value of $q$. Holding $M$ constant
and letting $q$ increase without bound, both parameters, $T_{i}$ and
$T_{b}$, decrease towards zero. Thus, as $q$ increases, the tachyonic
fluid domain shrinks down to a singularity and the entire
Schwarzschild $T$-domain is recovered.

\subsection{Regular Black Holes}
\qquad We next attempt to construct a reasonably general {\em
non-singular} $T$-domain solution. The philosophy here is to
study the (curvature) singularity properties of the space-time manifold.
As such, we employ a geometric method in solving the field equations since
we a priori have in mind certain geometric properties (namely, a
$T$-domain without singularities). The metric is prescribed here in
accordance with the non-singular properties dictated by the orthonormal
Riemann tensor. This, in turn, will dictate the
properties the matter field must possess in order to support non-singular
structure.

\qquad Solutions here are treated as
local and illustrate the general properties that solutions {\em must}
possess in order to be regular. We consider both the regular
Lorentzian and Instanton black hole. Some very interesting work
in this field regarding topology change may be found in \cite{ref:borde}.
Also, a regular black hole using a hypothesis of the existence of a
limiting curvature may be found \cite{ref:sing2}. There, a deSitter
universe is patched at the point of limiting curvature. In this
section we are loosely {\em using the same coordinates for all} $T <
2M$.
That is, the $T$ and $R$ coordinates are used for all domains
``inside'' the outer event horizon, even if it is an $R$-type or Euclidean
domain.

\qquad The first regular solution will correspond to an eternal $T$-sphere
with ``hard'' polytropic equation of state \cite{ref:weinberg}:
\begin{equation}
\rho=\kappa p , \label{eq:polytrope}
\end{equation}
with $\kappa$ a constant.
For such a solution consider the following metric for the matter domain:
\begin{equation}
ds^{2}=-\left(2CT^{2+x}-1\right)^{-1}\,dT^{2} +
\left(2CT^{2+x}-1\right)\,dR^{2} + T^{2}\,d\theta^{2} +
T^{2}\sin^{2}\theta\,d\phi^{2}, \label{eq:firstregmet}
\end{equation}
with $x > 0$ and $C > 0$.
Such a metric remains everywhere Lornetzian with the inner horizon located
at $T=T_{i}=1/(2C)^{\frac{1}{2+x}}$. It is easy to check that the energy
density
is proportional to both the parallel and transverse pressures.
Furthermore, in the $T$-domain, both pressures are negative or tensions.
This solution respects the weak energy condition since
\begin{equation}
\frac{\rho + p_{\shortparallel}}{|\rho|} \geq 0,
\end{equation}
as well as $\rho \geq 0$.

\qquad The non-zero orthonormal Riemann components are furnished by
\begin{subequations}
\begin{align}
R_{(TRTR)}=&-(2+x)(1+x)CT^{x},\\
R_{(T\theta T\theta)}=&-C(2+x)T^{x}\; , \\
R_{(R\theta R\theta})=&C(2+x)^{2}T^{x}\; , \\
R_{(\theta\phi\theta\phi)}=& 2CT^{x}\; ,
\end{align}
\end{subequations}
which are finite throughout the matter domain. In the special case $x=0$
this solution yields a matter domain of constant curvature similar to a
deSitter solution. This solution provides a simple, regular
polytropic $T$-domain model for a collapsed spherical star.

\qquad We next consider an example which is regular but may be {\em
Euclidean} in part of the manifold. It is unknown how physical this
situation may be. However, this case is worth dicscussing since it is
generally believed that at near the Planck scale, the concepts of space
and time may loose their meanings \cite{ref:wheel1}, \cite{ref:hart}.
Also, In astrophysical contexts, results in \cite{ref:das2} found
Euclidean instanton properties in the late collapse stages of a regular,
anisotropic star. As well, there is the well known Euclidean cosmological
instanton of Hawking and Turok \cite{ref:htinst} used to remove the big
bang singularity.

\qquad We wish to include a dependence on the interior radial coordinate,
$R$, to make the model more physical than $T$-spheres. Unfortunately,
adding
non-trivial $R$-dependence to the mass term usually leads to a singularity
at the inner horizon. Therefore, it is assumed that the mass term is
independent of $R$ and all $R$-dependence is incorporated through
$g_{RR}$.
The Riemann component $R_{(R\theta R\theta)}$ dictates that, in the
vicinity of $T=0$, $\alpha(T,R)$ behave as
\begin{equation}
\alf=\lambda(R)T^{y}+\xi(R), \label{eq:trueinstalf}
\end{equation}
with $y>2$.
Here $\lambda(R)$ and $\xi(R)$ are bounded, sufficiently differentiable
functions which are otherwise arbitrary and depend on the physical model.

\qquad Following the previous example, we set the line element as:
\begin{eqnarray}
ds^{2}&=&-\left(2CT^{2+x}-1\right)^{-1}\,dT^{2} +
\exp\left[\lambda(R)T^{y}+\xi(R)\right]\,dR^{2} \nonumber \\
& &+T^{2}\,d\theta^{2} +
T^{2}\sin^{2}\theta\,d\phi^{2}. \label{eq:trueinstmetric}
\end{eqnarray}
This solution is treated here as local within the vicinity of $T=0$ and
the inner horizon. It is useful so study such a solution since all
regular solutions subject to the restrictions above
must possess the form (\ref{eq:trueinstmetric}) if $T_{i}$ is sufficiently
``near'' $T=0$ (i.e. minimizing the Euclidean domain).

\qquad The metric (\ref{eq:trueinstmetric}) yields the orthonormal Riemann
components:
\begin{subequations}
\begin{align}
R_{(TRTR)}=&\frac{1}{2}\left[2CT^{2+x}-1\right] \left\{y\lambda(R)T^{y-1}
\left[-\frac{(2+x)T^{x+1}}{2CT^{2+x}-1}
\right.\right. \nonumber \\
&\left.\left.-\frac{1}{2}y\lambda(R)T^{y-1}
-(y-1)T^{-1}\right]\right\}, \\
R_{(T\theta T\theta)}=&-C(2+x)T^{x}, \\
R_{(R\theta R\theta)}=&\frac{1}{2}T^{y}\left(2CT^{+x}-T^{-2}\right), \\
R_{(\theta\phi\theta\phi)}=&2CT^{x}.
\end{align}
\end{subequations}
Note that this solution is regular both at $T=0$ and
$T=T_{i}=1/(2C)^{1/(2+x)}$.

\qquad Finally, the stress-energy tensor supporting this solution is
furnished by:
\begin{subequations}
\begin{align}
\Theta^{T}_{\;T}=&\frac{1}{8\pi}\left\{\lambda(R)yT^{y-2}-
2CT^{x}\left[1+\lambda(R)yT^{y}\right]\right\}, \\
\Theta^{R}_{\;R}=&-\frac{1}{4\pi}CT^{x}(3+x), \\
\Theta^{\theta}_{\;\theta}=&-\frac{1}{32\pi} \left\{ 4C(2+x)T^{x}
+2\lambda(R)yCT^{y+x}(2+2y+x) \right.\nonumber \\
 &\left.+\lambda(R)T^{y}
\left(2C\lambda(R)
T^{y+x}-\lambda(R)y^{2}T^{y-2}-2y^{2}T^{-2}\right)\right\}.
\end{align}
\end{subequations}
All physical quantities are therefore bounded at $T=0$ and $T=T_{i}$.

\section{Conclusion}
\qquad In this paper we have considered spherically symmetric matter
fields in the Schwarz\-schild $T$-domain. We developed the general
solution
as
well as some interesting particular solutions. The general solution
demonstrates that a locally measured {\em tension} is the source of the
invariant Schwarzschild mass which govern vacuum gravitational effects
felt by observers outside the black hole. Junction conditions which must
be met in order to patch matter solutions to the $T$-domain vacuum have
also
been addressed. The general approach laid out here also serves as a
reasonable starting point for future work in the $T$-domain.

\qquad Also considered was the $T$-domain analogue of constant density
stars. We found that the matter supporting such a structure is exotic
in the sense that principal stresses are tensions. Also, although the
energy density is positive, the matter field is locally tachyonic in
nature.

\qquad Finally, regular interiors were considered. There are two possible
scenarios for non-singular black holes addressed here. One is the presence
of a second horizon inside the black hole, restoring $R$-domain like
structure to the spacetime near $T=0$. The other is an ``instantonic''
phase which changes the spacetime signature to $+4$ as has been found in
some studies on gravitational collapse. There seems to be
no way to possess non-singular structure without resorting to one of the
above situations.

\newpage
\bibliographystyle{unsrt}

\end{document}